# TBDLNet: a network for classifying multidrug-resistant and drug-sensitive tuberculosis

Ziquan Zhu[1,#], Jing Tao[2,#], Shuihua Wang[1,3], Xin Zhang[2,*],Yudong Zhang[1,4,5,*],

1. School of Computing and Mathematical Sciences, University of Leicester, LE1 7RH, East Midlands, UK
2. Medical Imaging Department of Huai'an Fourth People's Hospital, 223002, Huai'an, China
3. Department of Biological Sciences, Xi'an Jiaotong-Liverpool University, Suzhou, Jiangsu 215123, China
4. School of Computer Science and Technology, Henan Polytechnic University, Jiaozuo, Henan 454000, P R China
5. Department of Information Systems, Faculty of Computing and Information Technology, King Abdulaziz University, Jeddah 21589, Saudi Arabia

# Ziquan Zhu & Jing Tao contributed equally to this paper and should be regarded as the co-first authors.

E-mail: Ziquan Zhu (zz257@le.ac.uk), Jing Tao (hasyyxk@163.com), Shuihua Wang (shuihuawang@ieee.org), Xin Zhang (973306782@qq.com), Yudong Zhang (yudongzhang@ieee.org)

* Co-corresponding authors

**Abstract:** This paper proposes applying a novel deep-learning model, TBDLNet, to recognize CT images to classify multidrug-resistant and drug-sensitive tuberculosis automatically. The pre-trained ResNet50 is selected to extract features. Three randomized neural networks are used to alleviate the overfitting problem. The ensemble of three RNNs is applied to boost the robustness via majority voting. The proposed model is evaluated by five-fold cross-validation. Five indexes are selected in this paper, which are accuracy, sensitivity, precision, F1-score, and specificity. The TBDLNet achieves 0.9822 accuracy, 0.9815 specificity, 0.9823 precision, 0.9829 sensitivity, and 0.9826 F1-score, respectively. The TBDLNet is suitable for classifying multidrug-resistant tuberculosis and drug-sensitive tuberculosis. It can detect multidrug-resistant pulmonary tuberculosis as early as possible, which helps to adjust the treatment plan in time and improve the treatment effect.

**Keywords:** multidrug-resistant tuberculosis; drug-sensitive tuberculosis; convolutional neural network; randomized neural network; ResNet50;

## 1. Introduction

Pulmonary tuberculosis is a chronic respiratory infectious disease caused by Mycobacterium tuberculosis [1]. Due to the lack of new anti-tuberculosis drugs, multidrug-resistant tuberculosis has spread, and multidrug-resistant pulmonary tuberculosis patients are characterized by strong infectivity, low cure rate, and high mortality [2]. The most commonly used methods and gold standards for clinical testing of drug resistance in pulmonary tuberculosis are sputum culture and drug sensitivity testing [3, 4]. Due to the time-consuming cultivation and drug-sensitivity testing of Mycobacterium tuberculosis, it is usually only considered whether it is drug-resistant after the failure of drug-sensitive tuberculosis treatment [5, 6]. Therefore, a large number of multidrug-resistant tuberculosis patients have not received a timely diagnosis.

Multidrug-resistant tuberculosis can be confirmed through sputum culture and sensitivity tests, bronchoalveolar lavage fluid, or lung tissue culture and sensitivity tests [7]. However, it may take days or even weeks to find acid-fast bacteria and determine drug resistance in sputum tests of tuberculosis patients. Multidrug-resistant tuberculosis has a long treatment time, poor treatment effect [8], and high mortality and treatment failure rate. Its treatment success rate is far lower than that of drug-sensitive tuberculosis. Although domestic and foreign scholars have conducted extensive research on the computer tomography (CT) manifestations of multidrug-resistant and drug-sensitive tuberculosis, there is no clear and unified differentiation standard. This requires the use of new vision to further distinguish between multidrug-resistant tuberculosis and drug sensitive tuberculosis, in order to provide new ideas and methods for their effective prevention and treatment.

CT is a commonly used diagnostic method for pulmonary tuberculosis, which can effectively diagnose and evaluate the efficacy of pulmonary lesions. Suppose medical personnel can use CT images to identify suspected drug-resistant tuberculosis patients early and conduct drug sensitivity tests in the laboratory in time. In that case, it can effectively shorten the diagnosis time of drug-resistant tuberculosis, adjust the treatment plan in time, and improve the treatment effect.

Convolutional neural network (CNN) proved its success in image tasks. Many researchers applied CNNs to the diagnosis of tuberculosis. Li, et al. [9] built an AECNN model to classify pulmonary tuberculosis based on CT images. The CNN was built to extract features. At the same time, the AutoEnconder was selected to extract unsupervised features. The proposed AECNN combined these features for the classification of pulmonary tuberculosis. This model could achieve 80.29% accuracy, 80.67% recall, and 80.42% F1. Meraj, et al. [10] applied four CNN models to detect pulmonary tuberculosis, which were ResNet50, VGG-16, GoogleNet, and VGG-19. Two public pulmonary tuberculosis datasets were used in this paper. The model with VGG-16 yielded the best accuracy, with 86.74% on the Shenzhen dataset and 77.14% on the Montgomery dataset.

Momeny, et al. [11] built a CNN model to classify tuberculosis based on microscopic images. The images were augmented based on square rough entropy. The Greedy AutoAugmentation model was used to alleviate the overfitting problem. In the new model, PReLU and dropout were implemented. Several different classifiers were selected for the classification. Finally, the model could get 93.4% accuracy. Wang, et al. [12] presented a GRAPNN network to classify pulmonary tuberculosis in CT images. The novel model named RAPNN was proposed to extract features from images. The graph convolutional network learned the relation-aware representation among the CT images. The GRAPNN achieved 94.88% accuracy, 94.65% sensitivity, 95.17% precision, 94.87% F1 score, and 95.12% specificity. Yi, et al. [13] built a new CNN model for the multi-classification of pulmonary diseases, which was named RED-CNN. The RED block was introduced to extract global image information, including the Res2Net, Double BlazeBlock, and ECA. The RED-CNN achieved 86.176% Jaccard scores, 91.796% recall, 91.796% accuracy, 91.892% F1 score, and 92.062% precision. Tian, et al. [14] produced a lightweight CNN model to classify pulmonary tuberculosis via CT images. The CNN module and Transformer module were combined to improve the classification performance. A shortcut was implemented in both modules to deal with the gradient divergence. This lightweight CNN model achieved 97.23% accuracy and 98.41% recall.

Recent research tried to classify multidrug-resistant tuberculosis and drug-sensitive tuberculosis. However, there are some deficiencies in these papers. Firstly, the performances of these papers are not ideal. Secondly, it is time-consuming to train the CNN models. To deal with these problems, we propose a novel deep-learning model to automatically classify multidrug-resistant tuberculosis and drug-sensitive tuberculosis, which is named TBDLNet. The proposed model aims to improve the accuracy of classifying tuberculosis and drug-sensitive tuberculosis in less time. Feature extraction was performed using pre-trained ResNet50. To address the overfitting problem, we employed three randomized neural networks (RNNs) for classification. Majority voting is used to improve the classification performance by using an ensemble of three RNNs.

This paper is structured as follows: The material is discussed in Section 2; Section 3 presents the methodology; Section 4 talks about the experiment results; and Section 5 concludes this paper.

## 2. Material

A retrospective analysis was conducted on 91 patients with multidrug-resistant pulmonary tuberculosis at the Fourth People's Hospital of Huai'an City from March 2018 to January 2023. The observation group consisted of 63 males and 28 females, aged 17-87 years, with an average age of 51.99+18.59 years. From March 2018 to January 2023, 7519 drug-sensitive pulmonary tuberculosis patients were randomly selected as a control group, consisting of 91 individuals, including 67 males and 24 females, aged 17-85 years, with an average age of 54.67+18.65 years. Inclusion criteria: (1) Positive sputum test for Mycobacterium tuberculosis, (2) Positive drug sensitivity test for multidrug-resistant patients, (3) Complete clinical diagnosis and treatment data, (4) Complete chest CT data. Exclusion criteria: patients with HIV infection, diabetes, and pneumoconiosis. These figures were conducted based on the Helsinki Declaration.

The Ethics Committee of the Fourth People's Hospital of Huai'an City approved the study without informed consent because it is a retrospective observational study, and all patient data is analyzed anonymously. All patients had positive sputum tests for Mycobacterium tuberculosis. A modified Roche culture medium was used to determine the sensitivity or resistance of the pathogen to drugs based on colony growth. The chest CT was scanned using Philips Ingenuity 64-row spiral CT, with the patient lying on their back and holding their breath from the lung tip to below the costophrenic angle after a deep inhalation. The scanning parameters were tube voltage 120KV, automatic tube current, layer thickness and pitch of 3mm, pitch of 1.0, and matrix 1024X1024.

The dataset used in this paper includes 14,125 images for the experiment. There are about 7,205 computerized tomography (CT) images of multidrug-resistant tuberculosis. The CT images of drug-sensitive tuberculosis are 6,920. The images of these two categories are given in Figure 1.

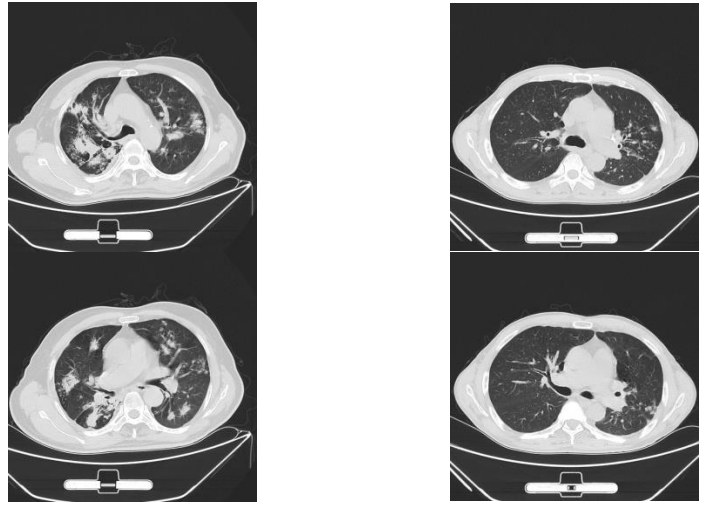

(a) Multidrug-resistant tuberculosis    (b) Drug-sensitive tuberculosis

Figure 1 Images of these two categories

## 3. Methodology

### 3.1 Backbone of Proposed TBDLNet

Recently, CNN models have been a promising choice for computer vision tasks. A sea of CNN models has been proposed and achieved outstanding results, such as AlexNet [15], VGG [16], MobileNet [17], and so on. ResNet [18] could be one of the most famous CNN models in recent years. The residual connection was proposed to approximate identity mappings by directly connecting two nonadjacent layers. The structure of the residual connection is presented in Figure 2.

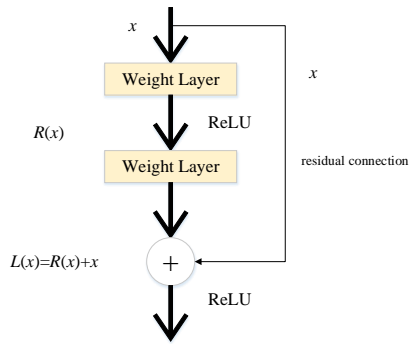

Figure 2 The residual connection

Given $x$ as the output of the last layer, training target $L(x)$ is denoted as. In the ResNet, the residual target $R(x)$ can be calculated as:

$$R(x) = L(x) - x. \qquad (1)$$

The training target can be presented as follows:

$$L(x) = R(x) + x. \qquad (2)$$

Many experiments have shown that applying CNN models to image classification tasks using transfer learning is better than training from scratch. The ability to extract high-level features has been captured by pre-trained CNN models, which can be transferred to downstream tasks. In this paper, the backbone model chosen is pre-trained ResNet50. Nevertheless, some modifications must be made due to the differences between these two datasets. The dataset used in this paper was created at the Fourth People's Hospital of Huai'an City from March 2018 to January 2023. There are 14,125 images for the experiment. There are about 7,205 computerized tomography (CT) images of multidrug-resistant tuberculosis. The CT images of drug-sensitive tuberculosis are 6,920. These modifications are shown in Figure 3. Six layers substituted the end three layers of the backbone model.

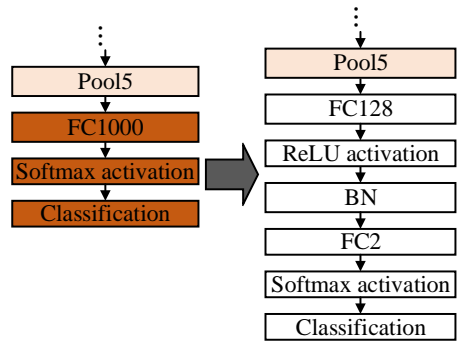

Figure 3 Modifications on the pre-trained ResNet50

**3.2 RNN for Classification**

CNN models have been proven to achieve great results for image tasks on big datasets. Nevertheless, when CNN models are implemented in small datasets, these models could meet the overfitting problem. The performance of CNN models in the small dataset is not very ideal. In this situation, the randomized neural network (RNN) is used for classification to alleviate the overfitting problem. The RNN has merely three layers: the input, hidden, and output layers. The simple structure of RNN helps cope with this problem. Three RNNs are added and replace the final five layers of the modified ResNet50, which are Schmidt neural network (SNN) [19], random vector functional link (RVFL) [20], and extreme learning machine (ELM) [21]. The frameworks of these three RNNs are shown in Figure 4.

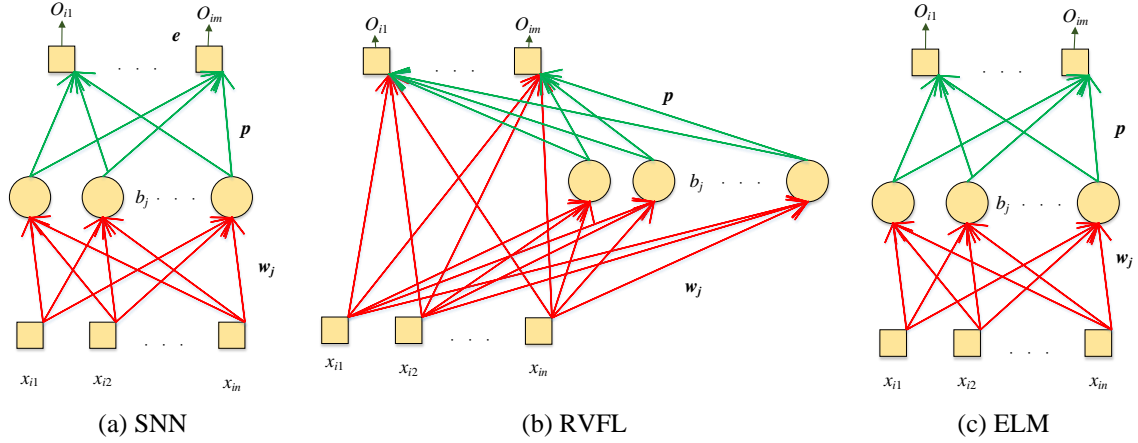

(a) SNN  (b) RVFL  (c) ELM

Figure 4 The frameworks of three RNNs

The frameworks of the three RNNs are different, but the training steps are the same. Given $N$ samples with $i$-th samples as $(x_i, y_i)$:

$$x_i = (x_{i1}, \ldots, x_{in})^T \in R^n, i = 1, \ldots, N, \tag{3}$$

$$y_i = (O_{i1}, \ldots, O_{im})^T \in R^m, i = 1, \ldots, N, \tag{4}$$

Firstly, the weight $w_j$ and bias $b_j$ are from the input layer to the hidden layer. The output of the hidden layer is calculated as follows.

For SNN:

$$\mathbf{M}_{\text{SNN}(i)} = \sum_{j=1}^{H} g(w_j x_i + b_j), i = 1, \ldots, N. \tag{5}$$

For RVFL:

$$\mathbf{M}_{\text{RVFL}(i)} = \sum_{j=1}^{H} g(w_j x_i + b_j), i = 1, \ldots, N. \tag{6}$$

For ELM:

$$\mathbf{M}_{\text{ELM}(i)} = \sum_{j=1}^{H} g(w_j x_i + b_j), i = 1, \ldots, N. \tag{7}$$

where $g()$ denotes the activation function and $H$ is the amount of hidden nodes.

Secondly, the output weight can be obtained by the pseudo-inverse.

For SNN:

$$(\mathbf{p}_{\text{SNN}}, \mathbf{e}_{\text{SNN}}) = \mathbf{M}_{\text{SNN}}^{+} \mathbf{Y}. \tag{8}$$

The concatenation of the input and the output of the hidden layer is used as the input for the output layer in RVFL:

$$\mathbf{C}_{\text{RVFL}(i)} = \text{concat}(\mathbf{X}, \mathbf{M}). \tag{9}$$

where $\mathbf{X} = (x_1, \ldots, x_N)^T$ is represented as the input feature set.

$$\mathbf{p_{RVFL}} = \mathbf{C_{RVFL}^+}\mathbf{Y}. \tag{10}$$

For ELM:

$$\mathbf{p_{ELM}} = \mathbf{M_{ELM}^+}\mathbf{Y}. \tag{11}$$

where $\mathbf{H_{net}^+}$ is the pseudo-inverse matrix of $\mathbf{H_{net}}$, $\mathbf{Y} = (\mathbf{y_1}, ..., \mathbf{y_N})^{\mathrm{T}}$ is the ground-truth label of the input, the output weight is shown as $\mathbf{p}$, and $\mathbf{e}$ is the bias of SNN.

**3.3 Ensemble of RNNs**

The randomly initialized parameters in the three RNNs have an excellent influence on reducing the training time. Meanwhile, these parameters are fixed and not updated through back-propagation, which benefits shortening time. However, ill-conditioned fixed weights and biases in RNNs could destroy the classification performance. Although individual networks could yield great classification performance, the ensemble of networks is more robust than individual networks. In this situation, three RNNs are ensembled to boost the results. Figure 5 gives the structure of the ensemble of RNNs.

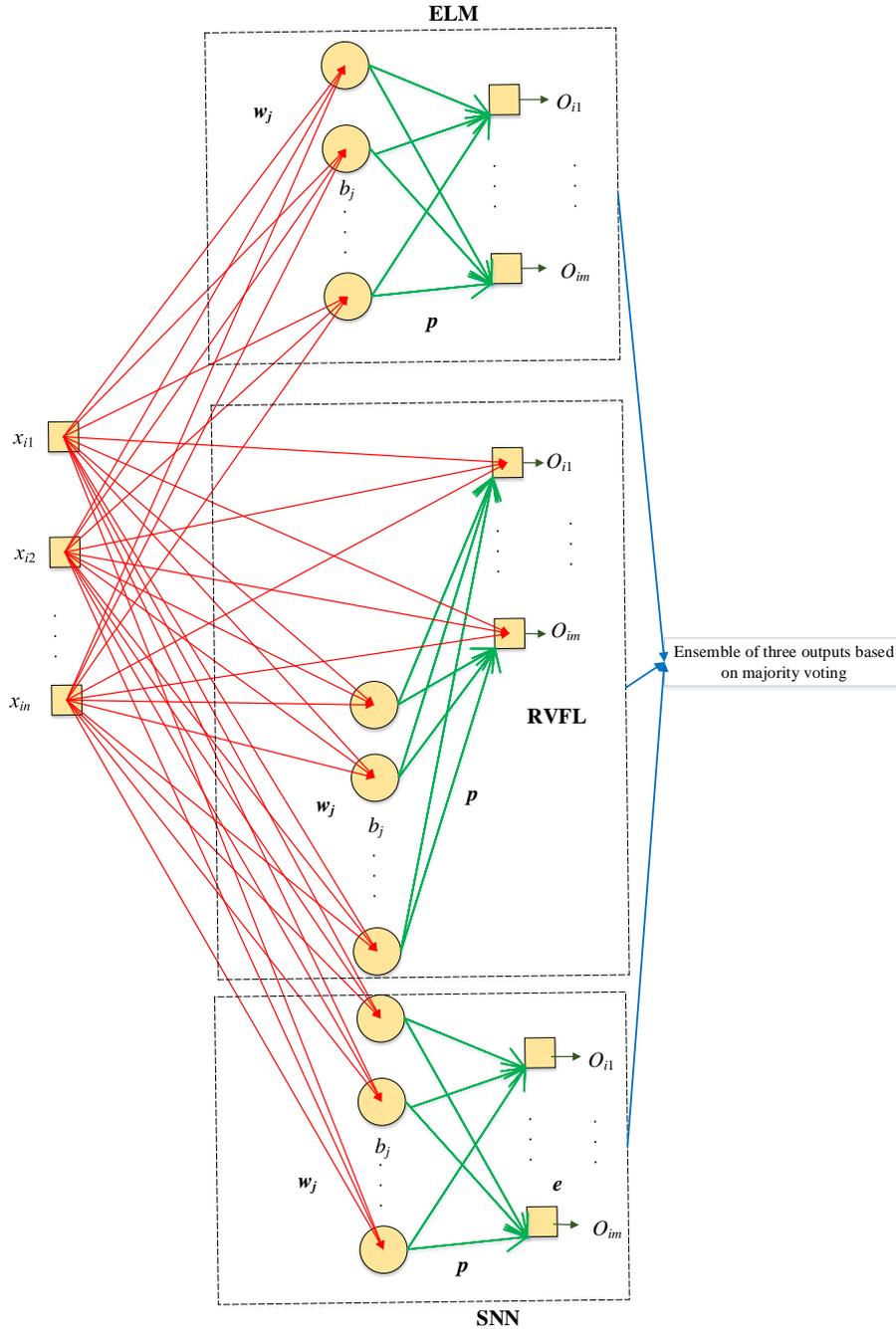

Figure 5 The structure of the ensemble of RNNs

**3.4 Proposed TBDLNet**

A new deep-learning model is proposed to classify multidrug-resistant and drug-sensitive tuberculosis, named TBDLNet automatically. In the proposed TBDLNet, the ResNet50 is implemented as the backbone model to extract features. The modifications are made on the pre-trained ResNet50 due to the differences between the dataset used in this paper and the ImageNet dataset. The training of CNN models is complex and time-consuming. Three RNNs are used and ensembled via majority voting. The structure of the TBDLNet is presented in Figure 6.

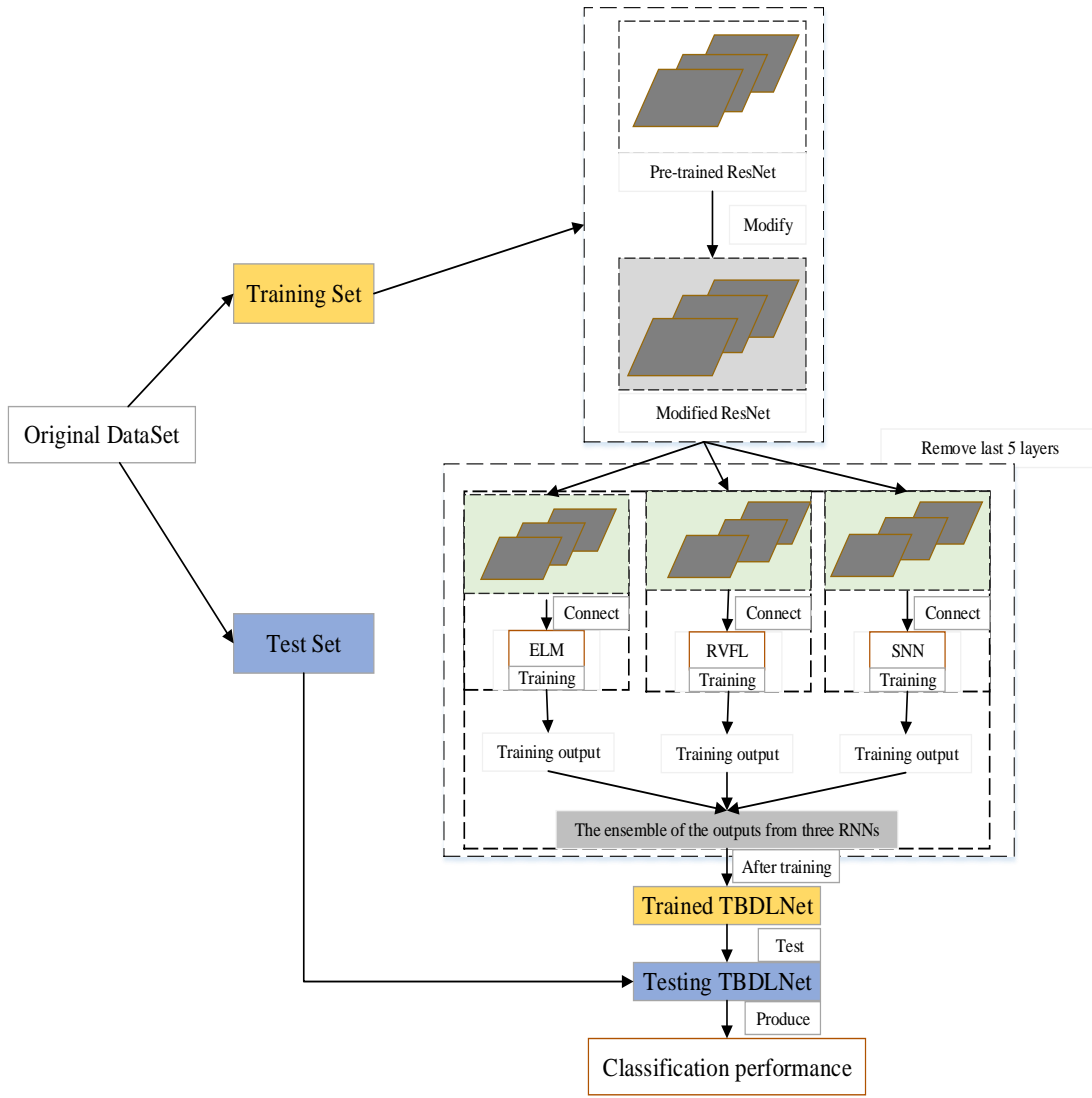

Figure 6  The framework of the proposed TBDLNet

## 3.5 Evaluation

The five-fold cross-validation to evaluate the TBDLNet. Five indexes are selected in this paper, which are accuracy, sensitivity, precision, F1-score, and specificity. These indexes are calculated as below:

$$\begin{cases} \text{Accracy} = \dfrac{TP + TN}{TP + TN + FP + FN} \\ \text{Sensitivity} = \dfrac{TP}{TP + FN} \\ \text{Precision} = \dfrac{TP}{TP + FP} \\ \text{Specificity} = \dfrac{TN}{TN + FP} \\ \text{F1} - \text{score} = \dfrac{2 \times TP}{2TP + FP + FN} \end{cases}, \qquad (12)$$

where TP, TN, FP, and FN are denoted as true positive, true negative, false positive, and false negative.

## 4. Experiment settings and results

### 4.1 Experiment Settings

The hyperparameter setting in this paper is presented in Table 1. For training the proposed model, the mini-batch size is set as ten due to the small size of the dataset used in this paper. The max-epoch is one by avoiding the overfitting problem. The initial learning rate is $10^{-4}$ based on experience. The number of hidden nodes of RNN is 400.

Table 1 The hyperparameter setting

| Hyper-parameter | Value |
|---|---|
| Mini-batch size | 10 |
| Max-epoch | 1 |
| Initial learning rate | $10^{-4}$ |
| $H$ | 400 |

### 4.2 The Performance of TBDLNet

The five-fold cross-validation is implemented to evaluate the TBDLNet. The final results of the proposed TBDLNet are presented in Table 2. It can be revealed that all results in each fold are greater than 0.97. The TBDLNet achieves 0.9822 accuracy, 0.9815 specificity, 0.9823 precision, 0.9829 sensitivity, and 0.9826 F1-score, respectively. In conclusion, the proposed model effectively classifies multidrug-resistant and drug-sensitive tuberculosis.

Table 2 The results of the TBDLNet

| | Accuracy | Specificity | Precision | Sensitivity | F1-score |
|---|---|---|---|---|---|
| Fold1 | 0.9830 | 0.9747 | 0.9761 | 0.9910 | 0.9835 |
| Fold2 | 0.9777 | 0.9790 | 0.9798 | 0.9764 | 0.9781 |
| Fold3 | 0.9798 | 0.9805 | 0.9812 | 0.9792 | 0.9802 |
| Fold4 | 0.9848 | 0.9834 | 0.9841 | 0.9861 | 0.9851 |
| Fold5 | 0.9858 | 0.9899 | 0.9902 | 0.9820 | 0.9861 |
| Average | 0.98222 | 0.9815 | 0.9823 | 0.9829 | 0.9826 |

### 4.3 Effects of RNN

Three RNNs are used to improve the classification performance. The results of these RNNs are demonstrated in Table 3. It is easy to see that the classification performance of ResNet50 with RNN is better than ResNet50. In conclusion, RNN can improve the classification performance for treating multidrug-resistant and drug-sensitive tuberculosis.

Table 3 The classification performance of these RNNs

|  | Accuracy | Specificity | Precision | Sensitivity | F1-score |
| --- | --- | --- | --- | --- | --- |
| ResNet50 | 0.9589 | 0.9776 | 0.9777 | 0.9410 | 0.9590 |
| ResNet50-RVFL | 0.9794 | 0.9827 | 0.9832 | 0.9764 | 0.9798 |
| ResNet50-SNN | 0.9706 | 0.9689 | 0.9702 | 0.9722 | 0.9712 |
| ResNet50-ELM | 0.9699 | 0.9747 | 0.9754 | 0.9653 | 0.9703 |

### 4.4 Effects of RNN Ensemble

RNN is considered an unstable network, even though it can perform well. We believe that the ensemble of RNNs could improve the robustness. The effects of the RNN ensemble are given in Table 4. Based on experiment results, the ensemble of RNNs could yield better results than individual RNNs. Therefore, the ensemble of RNN has sound effects on improving the robustness of the proposed network.

Table 4 The effects of RNN ensemble

|  | Accuracy | Specificity | Precision | Sensitivity | F1-score |
| --- | --- | --- | --- | --- | --- |
| ResNet50-RVFL | 0.9794 | 0.9827 | 0.9832 | 0.9764 | 0.9798 |
| ResNet50-SNN | 0.9706 | 0.9689 | 0.9702 | 0.9722 | 0.9712 |
| ResNet50-ELM | 0.9699 | 0.9747 | 0.9754 | 0.9653 | 0.9703 |
| TBDLNet | 0.98222 | 0.9815 | 0.9823 | 0.9829 | 0.9826 |

### 4.5 Effects of Backbone

We test our proposed model with different backbones: AlexNet, GoogleNet, MobileNet, ResNet18, and VGG. The comparison of different backbones is presented in Table 5. We can see that the ResNet50 can achieve the best results. Compared with other backbone models, the residual connection in the ResNet50 has a good effect on the classification performance. There are more layers in the ResNet50 than in the ResNet18. In this way, the deeper network could extract better features. The results of ResNet50 are better than the ResNet18.

Table 5 The comparison of different backbones

| Backbone | Accuracy | Specificity | Precision | Sensitivity | F1-score |
| --- | --- | --- | --- | --- | --- |
| AlexNet | 0.5812 | 0.7001 | 0.6186 | 0.4670 | 0.5322 |
| GoogleNet | 0.7632 | 0.8201 | 0.8039 | 0.7085 | 0.7532 |
| MobileNet | 0.9320 | 0.9364 | 0.9382 | 0.9278 | 0.9330 |
| ResNet18 | 0.9561 | 0.9682 | 0.9687 | 0.9445 | 0.9564 |
| VGG | 0.5094 | 0.0022 | 0.5098 | 0.9965 | 0.6745 |
| ResNet50 | 0.98222 | 0.9815 | 0.9823 | 0.9829 | 0.9826 |

### 4.6 Comparison with Other State-of-the-art Methods

We compare the TBDLNet with other state-of-the-art (SOTA) methods, which are GRAPNN [12], self-made CNN [22], Chi-squared and SVM [23], and TB-DRC-DSS [24]. The comparison performance is presented in Table 6. Based on the comparison performance, we can find that our proposed model can outperform other SOTA methods. It is easy to conclude that the proposed model is an efficient choice for classifying multidrug-resistant and drug-sensitive tuberculosis.

Table 6 The comparison results with other SOTA methods

| Method | Accuracy | Specificity | Precision | Sensitivity | F1-score |
|---|---|---|---|---|---|
| GRAPNN | 0.9488 | 0.9512 | 0.9517 | 0.9465 | 0.9487 |
| Self-made CNN | 0.5681 | - | - | - | - |
| Chi-squared and SVM | 0.7234 | - | - | - | - |
| TB-DRC-DSS | 0.9260 | - | - | - | 0.9090 |
| TBDLNet | 0.98222 | 0.9815 | 0.9823 | 0.9829 | 0.9826 |

### 5. Limitations and Discussion

In this paper, we propose a new model to classify multidrug-resistant tuberculosis and drug-sensitive tuberculosis. Our method significantly improves the classification performance of multidrug-resistant tuberculosis and drug-sensitive tuberculosis. The proposed model compares with other SOTA methods and yields the best results. Based on this great performance, the proposed model could be a good choice for daily diagnosis.

Even though the proposed model achieves great results, there are still some limitations to this model. Firstly, the proposed model is only applied to one data set. There are doubts about whether this model can still achieve outstanding results on other datasets. Secondly, this paper only pays attention to the classification. Segmentation and counting are very significant for diagnosis.

### 6. Conclusion

The most commonly used methods and gold standards for clinical testing of drug resistance in pulmonary tuberculosis are time-consuming. Therefore, a large number of multidrug-resistant tuberculosis patients have not received a timely diagnosis. A new network (TBDLNet) is proposed to classify multidrug-resistant and drug-sensitive tuberculosis. In the proposed model, the ResNet50 is chosen as the backbone model for the feature extraction. The modifications are made to the backbone model because of the differences between these two datasets. The RNN has merely three layers: the input, hidden, and output layers. The simple structure of RNN helps shorten training time and take up fewer computing resources. This proposed model has the following advantages compared to E-test, GeneXpert, and sequencing technology: low cost, fast speed, and high accuracy.

Three RNNs are implemented to boost the results: ELM, RVFL, and SNN. The randomly fixed parameters in the RNN make it unstable. The ensemble of three RNNs via majority voting is implemented to improve the robustness. The TBDLNet achieves 0.9822 accuracy, 0.9815 specificity, 0.9823 precision, 0.9829 sensitivity, and 0.9826 F1-score, respectively.

In the future, other technologies will be used to diagnose multidrug-resistant and drug-sensitive tuberculosis, such as Transformer, VIT, etc. CNN models could yield better performance on larger data sets. However, medical data sets are very small compared with other data sets such as ImageNet. Additionally, more information could be lost regarding the strong inductive biases in CNN models. In this way, Transformer and ViT could be choices to handle these problems. Moreover, the segmentation of tuberculosis is also an important topic. We should spend more time and energy on the segmentation of tuberculosis.

**Ethics statement**

The following information was supplied relating to ethical approvals: Medical Ethics Committee of Huai'an Fourth People's Hospital, China.

**Data availability statement**

Data will be available upon reasonable request.

**Conflict of interest statement**

The authors declare no conflict of interest.

**Funding**

The paper was partially supported by MRC, UK (MC_PC_17171); Royal Society, UK (RP202G0230); BHF, UK (AA/18/3/34220); Hope Foundation for Cancer Research, UK (RM60G0680); GCRF, UK (P202PF11); Sino-UK Industrial Fund, UK (RP202G0289); LIAS, UK (P202ED10, P202RE969); Data Science Enhancement Fund, UK (P202RE237); Fight for Sight, UK (24NN201); Sino-UK Education Fund, UK (OP202006); BBSRC, UK (RM32G0178B8); Huai'an Science and Technology Plan, China (HAB202329).

**Supplementary Material**

Figure 1 Images of these two categories
Figure 2 The residual connection
Figure 3 Modifications on the pre-trained ResNet50
Figure 4 The frameworks of three RNNs
Figure 5 The structure of the ensemble of RNNs
Figure 6 The framework of the proposed TBDLNet

Table 1 The hyperparameter setting
Table 2 The results of the TBDLNet
Table 3 The classification performance of these RNNs
Table 4 The effects of RNN ensemble
Table 5 The comparison of different backbones
Table 6 The comparison results with other SOTA methods